\documentclass[article,twocolumn,amsmath,amssymb,aps,prx]{revtex4-1}

\usepackage{graphicx}
\usepackage{dcolumn}
\usepackage{bm}
\usepackage{color}
\usepackage{xcolor}

\newcommand{\aver}[1]{\ensuremath{\langle {#1} \rangle}}

\begin{document}

\title{Quantum Effects in the Aubry Transition} 

\author{Pietro Maria Bonetti}%
\affiliation{ Max Planck Institute for Solid State Research,\\ Heisenbergstrasse 1, D-70569 Stuttgart, Germany}
\author{Andrea Rucci}
\author{Maria Luisa Chiofalo}
\email{maria.luisa.chiofalo@unipi.it}
\affiliation{ Dipartimento di Fisica “Enrico Fermi”, Universit\`a di Pisa and INFN,\\ Largo B. Pontecorvo 3, I-56127 Pisa, Italy}
\author{Vladan Vuleti\'c}
 \email{vuletic@mit.edu}
\affiliation{ Department of Physics, MIT-Harvard Center for Ultracold Atoms and Research Laboratory of Electronics, Massachusetts Institute of Technology,-
Cambridge, Massachusetts 02139, USA
 \\}

\begin{abstract}
The Aubry transition between sliding and pinned phases, driven by the competition between two incommensurate length scales, represents a paradigm that is applicable to a large variety of microscopically distinct systems. Despite previous theoretical studies, it remains an open question to what extent quantum effects modify the transition, or are experimentally observable. An experimental platform that can potentially reach the quantum regime has recently become available in the form of trapped laser-cooled ions subject to a periodic optical potential~\cite{Bylinskii2016}. Using Path-Integral Monte Carlo (PIMC) simulation methods, we analyze the impact of quantum tunneling on the sliding-to-pinned transition in this system, and determine the phase diagram in terms of incommensuration and potential strength.
We propose new signatures of the quantum Aubry transition that are robust against thermal and finite-size effects, and that can be observed in future experiments.

\end{abstract}

\pacs{Valid PACS appear here}
\maketitle
\section{Introduction}
An intriguing feature of quantum many-particle systems is that they may undergo phase transitions even at zero temperature, driven by the competition between kinetic and interaction energies across a critical point~\cite{Sachdev}. While the transition can be influenced by a variety of parameters such as the the dimensionality of the system, the strength and range of the interactions, or the amount of disorder, a particularly interesting situation arises when the phase transition is driven by the competition between two length scales. The latter mechanism gives rise to complex phenomena in a number of remarkably different physical environments, such as charge density waves~\cite{Gruner1988}, nanocontacts between solids~\cite{Mo2009}, dislocations in crystals~\cite{BraunBook}, adsorbed monolayers~\cite{Braun2006} in the form of noble gases on graphite substrates~\cite{ChaikinLubensky}, biomolecular transport~\cite{Bormuth2009}, emergence of chaotic structures in metal-insulator transitions in Peierls systems~\cite{Chiang1977} and spin glasses~\cite{Bak}, Josephson junctions~\cite{Kurten2012}, quantum pinning in strongly-interacting bosonic fluids~\cite{Haller} and Meissner-to-vortex transition in bosonic ladders~\cite{Atala,Orignac}. This non-exhaustive list illustrates how the concept of competing length scales underlies a wide range of phenomena from classical to quantum, with the latter prominently emerging in reduced dimensions, where quantum fluctuations and correlations are greatly enhanced.

Various interesting behaviors can then emerge, depending on whether the two lengths are commensurate (C), i.e. in a rational ratio, or incommensurate (IC, irrational ratio), and how the incommensuration is accommodated by the system. Incommensurations can, for example, float on the commensurate phase~\cite{Consoli2000}, or instead occur as interstitial phases separating 
commensurate phases~\cite{ChaikinLubensky}.
In general, a transition may occur from a pinned to a floating, translationally invariant-phase.
The incommensurate phase can be associated with a manifestation of translational invariance in the form a gapless excitation called phason, corresponding to superlubric sliding. In the commensurate phase, translational invariance is broken, and a gap appears in the excitation spectrum that corresponds to the energy required for the classical particle to climb the barrier. Above the transition point, no particles are found at the lattice maxima, as elucidated by Aubry~\cite{Aubry1983}.
Defects in the pinned phase can be in the form of walls, dislocations, or vortices~\cite{Bak}.

The paradigmatic model to describe this general situation is a one-dimensional (1D) chain of particles connected by harmonic springs, separated at equilibrium by a distance $d$, and subjected to a substrate lattice with period $a$. This model has been independently proposed by Frenkel and Kontorova ~\cite{FK1938}, and Frank and Van der Merwe~\cite{FVdM} (FKVdM). In the FKVdM model, the competition is between the elastic potential, which favors a periodic structure with period $d$, and the lattice potential which tries to lock the particle positions at integer multiples of $a$. As a result, in the limit of small lattice potential $V$, the particles float on the lattice almost independently of the ratio $w\equiv d/a$ in the IC phase, while in the C phase, the particles localize near the lattice minima in a commensurate structure with an average spacing that is a rational multiple of $a$. The transition can be understood in more detail by introducing the average spacing $\overline{d}$ between the atoms in the chain and the winding number $\tilde{w}=\overline{d}/a$. 
At $V=0$, $\tilde{w}=w$, and the particles float. A finite lattice depth acts to localize the particles. Non-rational values of $w$ are accommodated by an energy increase that maintains the winding number $\tilde{w}=w$, until there is sufficient energy to form a discommensuration. The C-IC transition predicted by Aubry~\cite{Aubry1983} sets in as a second-order transition, and the winding $\tilde{w}$ jumps up to the next rational number. Below a critical lattice strength $V_c$, $\tilde{w}$ is irrational in finite intervals of $w$, forming a Cantor set with non-integral fractal dimension~\cite{ChaikinLubensky}, the interval size reducing to zero measure above threshold $V>V_c$. The critical lattice depth $V_c$ depends on the irrational value of the length ratio $d/a$, taking on its largest value at the golden ratio $(\sqrt{5}-1)/2$. Notably, the transition survives even when the number of particles is finite~\cite{Sharma1984}.

The quantum version of the FKVdM model has also been investigated~\cite{Borgonovi}.
Quantum Monte Carlo (QMC) studies for up to 144 particles have revealed that quantum mechanics tends to smooth the onset of the Aubry transition~\cite{Hu2000}. Indicators of the transition have been defined in close analogy to the classical transition and include the hull function, the position variance of the particles, and the density-density correlation function. An effective Planck constant
to measure the degree of quantum behavior has been identified~\cite{Hu2000,Zhirov}.  Correlations in the C phase have been investigated via Density-Matrix Renormalization Group (DMRG)~\cite{Ma} and Path-Integral Molecular Dynamics methods~\cite{Krajewski}, while the IC phase has been studied by DMRG~\cite{Hu}. Excitations across the transition have been discussed in terms of a pinned instanton glass turning into a sliding phonon gas~\cite{Zhirov}. The effect of long-range interactions has been investigated in Ref.~\cite{Pokrovsky}. The formation of kinks, i.e. topological solitons, and kinks-antikinks pairs within a nonlocal FKVdM model in the presence of long-range, power-law interactions, has been characterized~\cite{Braun90}, also accounting for finite-size effects~\cite{Landa20}. The relevance of the FKVdM-model to structural zigzag instabilities of ion strings~\cite{Silvi} in optical resonators has also been pointed out~\cite{Cormick,Gangloff20,Fogarty}. 
Besides simulations, theoretical methods include the possibility of describing the C-IC transition in terms of an Ising model, as in the seminal work by P. Bak and R. Bruinsma~\cite{BakBruinsma}.

Notwithstanding this wide range of previous studies, a number of crucial questions regarding the quantum Aubry transition  remain open, such as the universality class of the transition, the relation between the C-IC transition and the paradigm of many-body localization, the degree of universality for different types of interactions,
and the general experimental feasibility of observing quantum effects in the FKVdM model in realistic systems of finite size and temperature.

In a recent experiment using a small chain of trapped Yb$^+$ ions, Bylinskii {\it et al.} have observed an Aubry-like transition ~\cite{Bylinskii2016} at low temperatures where quantum effects may come into play by studying friction, and investigated the dependence on temperature~\cite{Gangloff2015} and commensurability~\cite{Bylinskii2015}. In this finite-size system, the competition between the two different length scales, given by the period of the applied optical potential and the average interparticle spacing of the trapped particles, drives a transition between pinned and sliding arrangements. Using suitably devised atom-by-atom control and observation techniques, the transition from superlubricity to stick-slip has been investigated as a function of the height of the optical potential and the commensurability of the two length scales ~\cite{Gangloff20}, thus demonstrating a versatile and controllable platform to explore the building blocks of friction at a nanoscopic level~\cite{Benassi11,Mandelli13,Puttivarasin11,GarciaMata07} over a wide temperature range. In a second ion-trap experiment in two dimensions (2D)~\cite{Kiethe17}, the underlying nanofriction and transport processes have been investigated with atomic resolution using spectroscopic measurements, enabling the observation of a soft vibrational mode through the transition.

In this work, we investigate how to unequivocally identify quantum effects in the Aubry transition, and whether they can be observed in present finite-size ion-chain experiments that provide only an approximation to the FKVdM model. To this aim, we perform PIMC simulations reproducing the finite temperature and other conditions of the experiment \cite{Bylinskii2015}, and propose new observables -- never considered so far -- as consistent indicators of emerging quantum behavior. We show that the quantum effects are sufficiently robust with respect to thermal fluctuations at the temperatures that have already been experimentally achieved, and that they can also be readily distinguished from finite-size effects already in the relatively small systems that are currently available.

The paper is organized as follows: In Sec.~\ref{sec:System} we present the system and the discuss the governing parameters. After introducing the methodological tools in Sec.~\ref{sec:Method},  we discuss in Sec.~\ref{sec:Results} the simulation results and their theoretical implications, devoting Sec.~\ref{sec:Experiment} to a discussion how to observe quantum signatures in the actual experiment. We finally present conclusive remarks in Sec.~\ref{sec:Conclusions}.\\

\section{The System}\label{sec:System}
\begin{figure}
    \centering
    \includegraphics[width =  1.\linewidth]{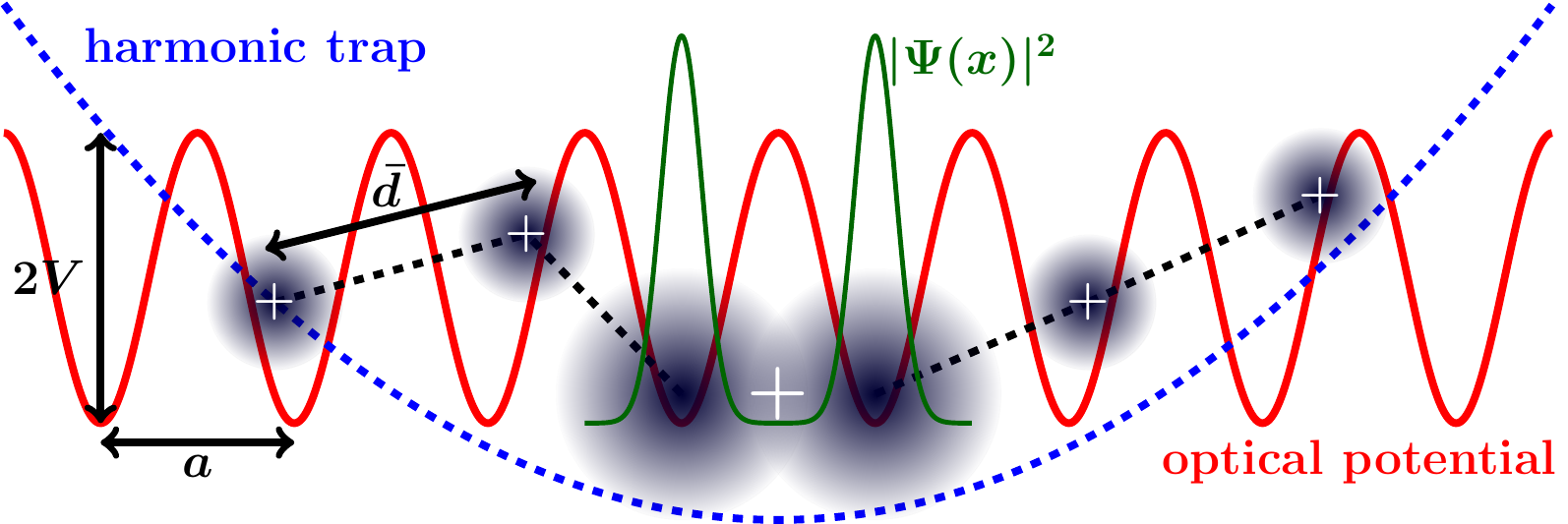}
    \caption{Concept for experimental realization of the quantum Aubry transition with laser cooled trapped ions subject to an optical-lattice potential following Ref. \cite{Bylinskii2016}. $N$ ions  with charge $+|e|$ are separated on average by a distance $\overline{d}$ that results from the combined action of the mutual Coulomb repulsion and the external harmonic confinement. Overlaid is a periodic optical lattice with lattice constant $a$ and height $V$.
    The length scales $a$ and $\overline{d}$ are in an irrational or incommensurate ratio.
    The ions may float on the periodic lattice in the incommensurate phase for $V$ below a critical value $V_c$, or pin into commensurate pieces separated by incommensurations for $V>V_c$. Quantum effects are driven by tunnelling between two neighboring lattice sites. Here, the quantum probability density $|\Psi(x)|^2$ for the central particle is sketched, as calculated with a PIMC simulation.}
    \label{fig1}
\end{figure}

The system concept is outlined in Fig.~\ref{fig1}. 
Laser-cooled, positively-charged ions arranged in a 1D chain interact with each other via Coulomb repulsive forces while being overall confined 
in a harmonic potential characterized by a trap frequency $\omega_0$. The combined action of the Coulomb repulsion and the harmonic confinement sets the ion arrangement with an average interparticle distance $\overline{d}$. An optical lattice provides the static periodic potential with height $V$ and period $a$, with the two length scales $\overline{d}$ and $a$ being in a (quasi-) incommensurate ratio. Notice that $\overline{d}$ is calculated in the absence of the lattice, i.e. for $V=0$.  Depending on whether the lattice-potential height $V$ is below or above a critical value $V_c$, the ions either float on the lattice in a superlubric phase, or are pinned into the lattice wells, forming commensurate regions separated by incommensurate ones.  

We thus consider the following Hamiltonian, in cgs units, precisely reproducing the experimental realization from Ref.~\cite{Bylinskii2016}: 
\begin{equation}
\begin{split} 
\mathcal{H}=&\sum_{i=-(N-1)/2}^{(N-1)/2} \Bigg\{ \frac{p_i^2}{2m}+\frac{1}{2}m\omega_0^2 \left(x_i-\frac{a}{2\pi}\Phi\right)^2+\\
+&V \bigg[1+\cos\left(\frac{2\pi}{a} x_i\right)\bigg]\Bigg\}
+\sum_{i,j>i}^{N}\frac{e^2}{\left | x_i-x_j\right|} .
\end{split}
\label{eq:H}
\end{equation}
The first term represents the kinetic energy of the $N$ ions having mass $m$, with $N$ being an odd number, the second term the harmonic confining potential with oscillator frequency $\omega_0$, while the third accounts for the the periodic potential with height $V$ and period $a$, with a a possible shift $\Phi$ between the minima of the harmonic trap and of the optical potential. 
The last term represents the Coulomb interactions between the $N$ ions, each possessing a charge $+|e|$. Whenever not explicitly specified, the value $\Phi=0$ is considered. 

The system is thus characterized by two different length scales, namely $\overline{d}$ and $a$. (While the spacing $d_i\equiv|x_{i+1}-x_i|$ between neighboring ions in the chain is non-uniform, the ratio $\overline{d}/a$ is sufficiently large that approximately the same incommensurability $(d_i/a \mod 1)$ for all ions can be achieved by slight adjustment of the harmonic-trap potential~\cite{Bylinskii2016}.) The combined action of the Coulomb and harmonic-oscillator potentials acts as a spring-like internal force with effective spring constant  $\gamma$ after linearization. As a result, $\overline{d}$ and $a$ define together the elastic energy 
\begin{equation}
E_e=\gamma \left(\frac{a}{2\pi}\right)^2;\qquad \gamma=\frac{2e^2}{\overline{d}^3},
\label{eq:ElasticE}    
\end{equation}
which will be used as our energy unit $E_e$ from now on. Notice that $E_e$ depends on the harmonic-oscillator frequency $\omega_0$ through the average ion spacing $\overline{d}$. In addition to the effective spring energy arising from the Coulomb potential, there are external forces due to the periodic potential, characterized by $K={V}/{E_e}$ in dimensionless units, and due to the harmonic trap confinement. To further account for the two length scales $\overline{d}$ and $a$, we introduce a third parameter $\Delta$, quantifying the incommensuration of the system via 
\begin{equation}
\displaystyle{\Delta=1-\max_{\theta} \left \{ \frac{1}{N}\sum_{j=-(N-1)/2}^{(N-1)/2}\cos\left(\frac{2\pi}{a}\langle x_j \rangle+\theta\right)\right\}},
\label{eq:Delta}    
\end{equation}
where $\langle x_j \rangle$ is the average position of the $j$-th particle measured at vanishing periodic potential $K=0$, and depends on the parameters of the harmonic potential. In this manner, $\Delta=0$ corresponds to a maximally commensurate system, while $\Delta=1$ corresponds to a maximally incommensurate situation. Overall, we see that the harmonic confinement plays three different roles: it stabilizes the internal repulsive forces, it produces a non-homogeneity in the external forces, and it may be used to adjust the commensurability of the system. 

We choose the same physical parameters to those typical in the experiment~\cite{Bylinskii2015}, so that  $k_BT\simeq0.01\, \hbar\omega_0$, i.e. the lowest reachable temperature, and $a=185$~nm. Simulations at different temperatures are performed in order to check the robustness of quantum effects in the presence thermal fluctuations. The other parameters are varied in the following ranges: $\omega_0/(2\pi)$=(2.09-2.11)~MHz, and $V/k_B$=(0-1.60)~mK. These ranges correspond to $\overline{d}/a\simeq$ $10^3$. Notice that despite this large ratio, the parameter that drives the system across the transition is the quantity $(\overline{d}/a \mod 1)$, so that even small variations in $\overline{d}$ can significantly modify the system behaviour.

\section{The Method}\label{sec:Method}
In order to determine the ground state of Hamiltonian \eqref{eq:H}, we resort to a Path Integral Monte-Carlo
(PIMC) simulation at finite temperature $T$~\cite{Rothe}, where the path integrals are computed within a Markov-Chain Monte-Carlo (MCMC) simulation~\cite{Rothe}. Finite temperature enters after executing a Wick rotation ($t\rightarrow i\tau$) in the classical action. Thus, time evolution occurs in an imaginary time 
axis with length $\hbar/k_BT$, and periodic boundary conditions. In this manner, each particle is characterized by a loop in imaginary time. The configurations, i.e.  the collection of all particle paths, follow a probability distribution given by the Boltzmann weight. Configurations are then sampled according to the distribution, and the observables are calculated on that sample.
In performing this operation, we take special care of autocorrelation effects of next-observable measurements and of time discretization~\cite{SoM}. In particular, all the presented data result from having performed simulations for three different values of the time-step $\tau$, extrapolating the limit $\tau\rightarrow 0$ from a linear fit. Finally, special care has also been taken in order to efficiently deal with tunneling processes arising when the minimum of the harmonic trap coincides with a maximum of the optical potential (i.e. when $\Phi=2n\pi$ in eq. \eqref{eq:H}). For more details on this technical choice, we refer to the Supplemental Material~\cite{SoM}.\\

\section{Results}\label{sec:Results}
\subsection{Emergence of quantum effects and bimodal probabilities}
In order to mark the onset of the transition, an order parameter needs to be identified. In particular, we seek a quantitative parameter that indicates whether the single particle wave function has a gaussian or a bimodal shape. To this end, we consider the Binder cumulant (BC), defined as:
\begin{equation}
B_j\equiv 1-\frac{\langle (x_j-\overline{x}_j)^4\rangle}{3\langle(x_j-\overline{x}_j)^2\rangle^2},
\label{eq:Binder}
\end{equation}
where $\overline{x}_j=\langle x_j\rangle$ and the averages are to be calculated on the single-particle probability distribution $\varrho_j(x)$. In our case, this is the probability density of a given ion, derived from the many-particle wavefunction that is traced over the other particles, the total number being $N=5$ as in the experiment~\cite{Bylinskii2015}. In fact, the BC translates a qualitative shape of the wavefunction into a quantitative indicator. A gaussian probability distribution is characterized by $B=0$, while a bimodal distribution composed  of two symmetric Dirac delta functions yields $B=2/3$. Therefore, the BC is a quantity that can distinguish between spatial
probability distributions with one or two peaks, thereby providing a
clear signature of the emergence of quantum effects driven by tunneling.

\begin{figure}
    \centering
    \includegraphics[width =  1.\linewidth]{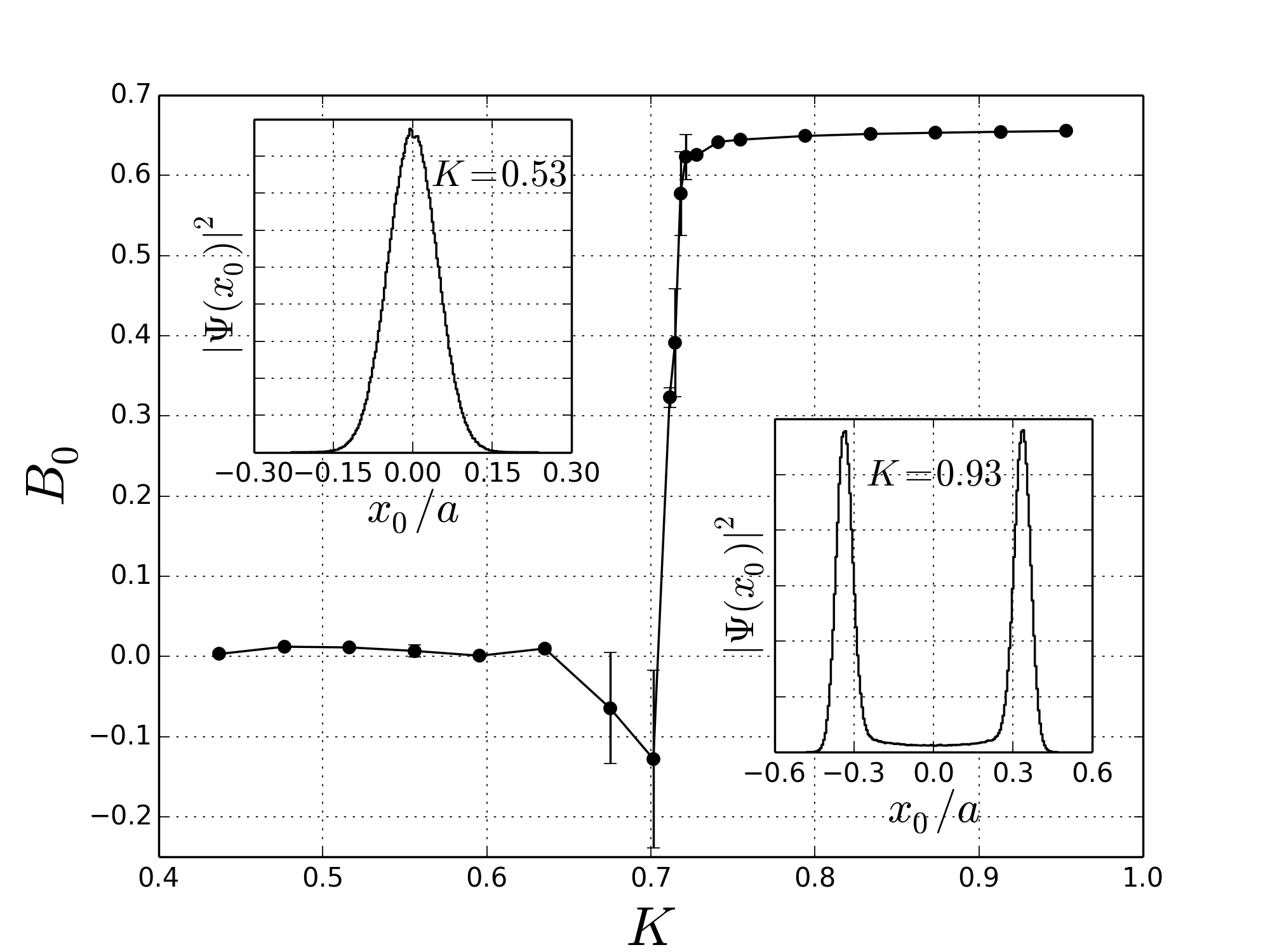}
    \caption{Characterizing quantum effects via the Binder cumulant $B_0$. Behaviour of $B_0$ (eq. \eqref{eq:Binder}) for the probability distribution of the central particle as a function of the scaled optical lattice depth $K$ in a fully commensurate system. Notice the sharp increase of $B_0$ at the value $K\simeq K_c \simeq 0.7$. 
    For low values of the potential, $B_0\simeq 0$, indicating a gaussian-like distribution, while for large $K$ values we find $B_0\simeq 2/3$, characteristic of a bimodal distribution. Upper left inset: gaussian-like probability distribution (modulus squared of the single particle wave function) for $K<K_c$. Lower right inset: bimodal distribution for $K>K_c$.
    }
    \label{fig2}
\end{figure}

We first test the BC on the simplest commensurate situation. To this aim, we calculate the BC while varying $K$, and first analyze the behavior of the central particle, as shown in Fig.~\ref{fig2}. The corresponding value $B_0$ of the BC is observed to sharply increase at $K_c\simeq 0.7$ towards the limiting value of $2/3$. As long as the system is in the sliding phase for $K<K_c$, the wavefunction also keeps the single-mode structure depicted in the top-left inset. In the opposite $K>K_c$ regime of the pinned phase, the bimodal structure in the bottom-right inset is observed, as a signature of quantum tunneling leading to the localization in two wells. 

Zooming in to the transition point, we  notice a region where $B<0$. We find that when approaching the transition point, the tails of the probability distribution begin to extend, so that the fourth moment exceeds the second moment by more than a factor of 3. However, near the transition, the PIMC convergence slows down, and statistical errors are larger, preventing us from further exploring the details of the probability distribution.

\subsection{The effects of incommensuration}
\begin{figure}
    \centering
   \includegraphics[width =  1.\linewidth]{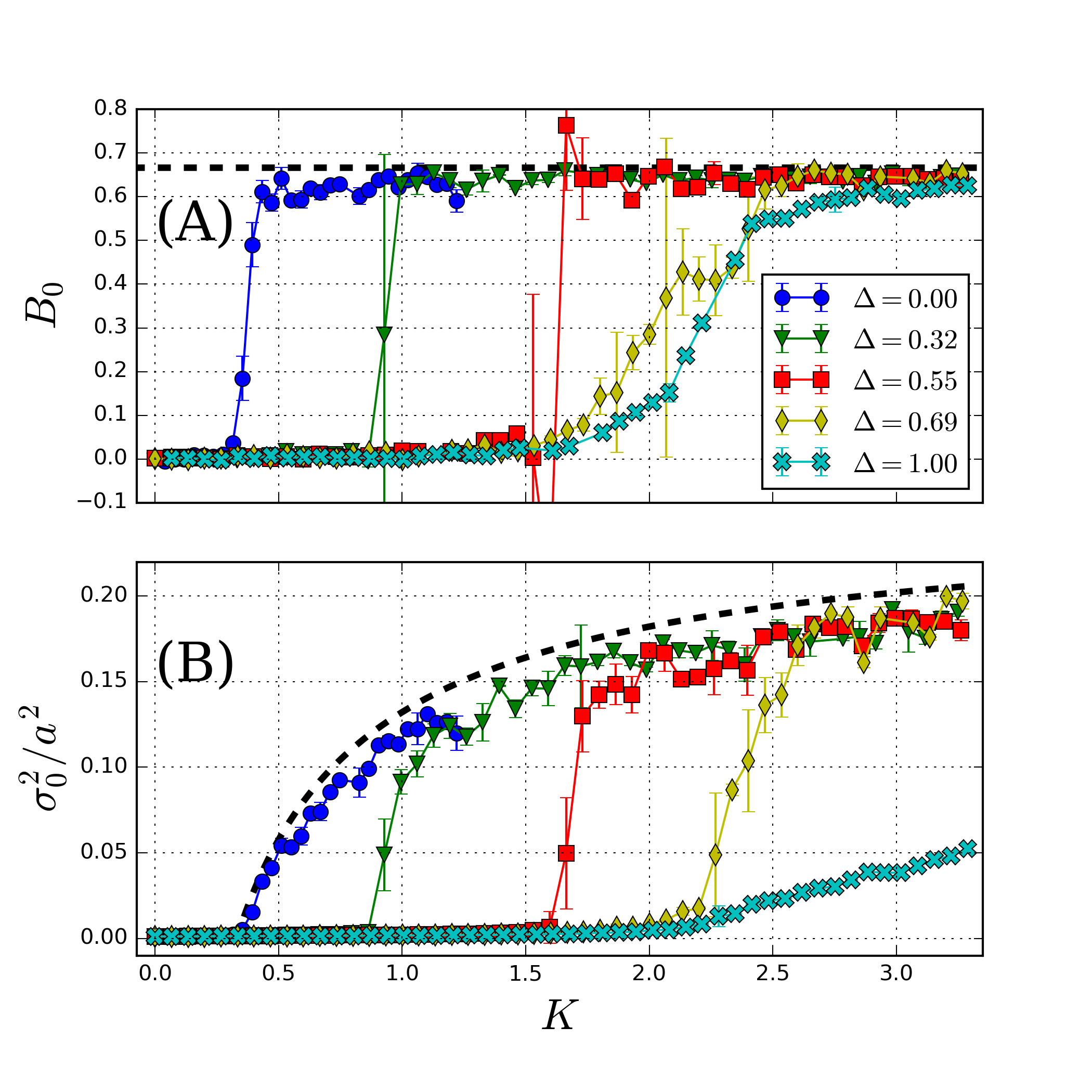}
    \caption{Incommensuration effects. (A) Behaviour of the Binder cumulant of the central particle, $B_0$, as a function of potential depth $K$ for different incommensuration parameter values $\Delta$ (eq. \eqref{eq:Delta}). $\Delta=0$ corresponds to a maximally commensurate and $\Delta=1$ to a maximally incommensurate system. The potential configuration is the one in which the minimum of the trap coincides with an optical-lattice maximum. The dashed line represents the value $2/3$ expected for a bimodal distribution.  
    (B) Variance $\sigma_0^2$ of the central particle under the same conditions as for (A). As for the Binder cumulant, we notice that larger incommensuration $\Delta$ moves the critical potential depth $K_c$ to larger values and softens the transition. The dashed line represents the variance for the classical position of the central particle as described in the text. It is interesting how for all incommensurations, with the exception of $\Delta=1$, the variance approaches this universal classical behavior.}
\label{fig3}
\end{figure}

Having established that the BC provides a quantitative signature of the transition, we now turn to explore the effects of  incommensuration. Fig.~\ref{fig3} displays the behaviour of the BC of the central particle, $B_0$, as a function of the lattice potential strength $K$ for different values of the incommensuration parameter $\Delta$. We notice that a larger $\Delta$ increases the critical pinning potential $K_c$, i.e. the more incommensurate the system, the more it resists pinning. At the same time, the transition becomes less sharp compared to the corresponding commensurate case in Fig.~\ref{fig2}. In fact, in the commensurate system the springs have no effect (they are neither compressed nor stretched), and all the particles move in unison but independently, reducing the transition to that of a single-particle one. On the other hand, under incommensurate conditions ($\Delta\neq 0$), it may happen that some of the particles do not undergo the pinning transition because they are at the minima of the optical lattice, while others are near the maxima. This requires larger values $K$ of the potential to complete the transition. Thus the transition region is widened as each particle becomes pinned at a different value of $K_c$. The smoothing is seen to persist in a ring geometry without the external confinement~\cite{SoM}, and in the maximally incommensurate case has been observed also in ~\cite{Hu2000}. The same smoothing does not happen in the classical Aubry model, where the transition is sharp and simultaneous for all particles \cite{Aubry1983}.

Similar conclusions can be drawn from the position variance $\sigma_0^2=\aver{x_0^2}-\aver{x_0}^2$ of the central particle for $\Phi=0$. This variance is displayed in Fig.~\ref{fig3}B as a function of $K$ for different incommensuration values $\Delta$. Above a critical value $K_c$ that agrees with that inferred from the BC $B_0$, the variance $\sigma_0^2$ increases. As already pointed out in~\cite{Hu2000}, $\sigma_0^2$ is a useful observable, though it cannot be considered, strictly speaking, an order parameter. We can compare the simulations of \emph{Hu et al.}, performed with up to 144 particles connected by springs~\cite{Hu2000}, with our 5-particle system interacting via Coulomb forces, and find that the steepness of the transition is similar for comparable parameter values. We may thus infer that, at least for small incommensurations, our steep signature of the transition is robust against the effects of finite system size. Second, as with the BC behavior, larger values of the incommensuration parameter $\Delta$ move the critical $K_c$ to larger values, and broaden the transition. 

However, two differences are evident when comparing the BC ($B_0$ in Fig. \ref{fig3}A) and the variance $\sigma_0^2$ of the central particle (Fig. \ref{fig3}B). First, in the maximally incommensurate case with $\Delta=1$, the variance never reaches the limiting value, while the BC does.
Second, for $\sigma_0^2$ the steepest behaviour occurs for intermediate incommensuration, in contrast to the behaviour for $B_0$.

The origin of this effect can be understood by calculating the classical position distribution of a single particle as a function of $K$. To this aim, we consider the many-particle potential $V\big(x_{-({N-1})/{2}},...,x_{({N-1})/{2}}\big)$, composed of external confinement, optical lattice, and Coulomb interactions, and minimize it over the variable set $\big\{x_{-({N-1})/{2}},...,x_{({N-1})/{2}}\big\}$, obtaining the values $\overline{x}_i$ of the particle positions that minimize $V$. Finally, we retain only the central particle position $\overline{x}_{0}$. In principle, the positions of the minima depend on the harmonic trap frequency, but we find that in the asymptotic limit of large $K$ this is not the case. The classical position of the central particle, calculated by the illustrated method, is in fact represented by the dashed line in Fig.~\ref{fig3}, which evidently fits the numerical data in the $K\to\infty$ limit, without any fitting parameter. Indeed, when $K$ is increased above the critical value $K_c$, the central particle wavefunction resembles more and more a double delta distribution peaked at the classical minima of the many-particle potential $\pm \overline{x}_0$, i.e. $\Psi(x)\simeq 1/2[\delta(x-\overline{x}_0)+\delta(x+\overline{x}_0)]$, whose variance is indeed $\overline{x}_0^2$. In contrast, in the maximally incommensurate case ($\Delta=1$), the central particle variance is evidently far below this asymptotic value because the lack of localization of the other particles above the threshold $K_c$ broadens the central-particle wavefunction around the two peaks.

\subsection{The phase diagram}

\begin{figure}
    \centering
   \includegraphics[width =  1.\linewidth]{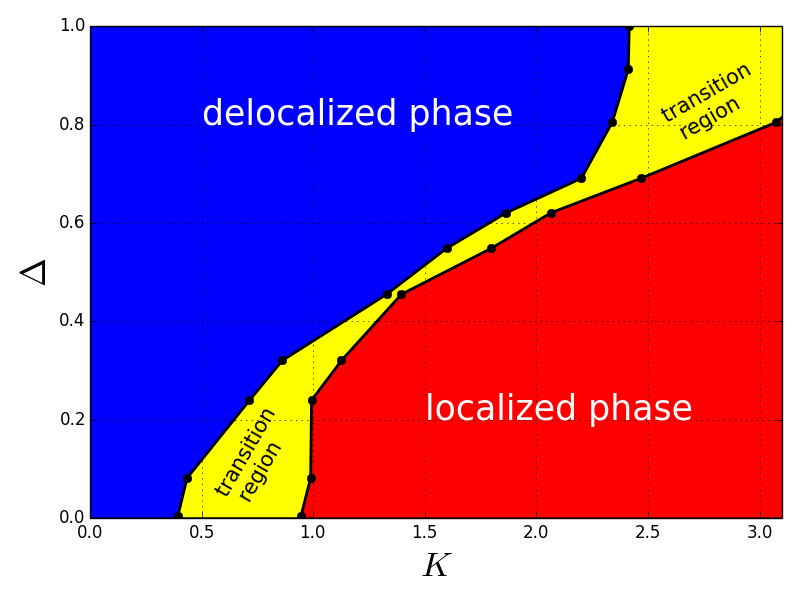}
    \caption{Phase diagram of the Aubry transition. The variance of the central particle is displayed in $\Delta$-$K$ space. Red region: localized phase. Blue region: delocalized phase. Yellow region: intermediate phase. For the sake of definiteness, we have quantitatively defined them after choosing the following thresholds: $\sigma_0^2<0.02a^2$, $a^2<\sigma_0^2<0.11a^2$, and $\sigma_0^2>0.11 a^2$ for the delocalized, intermediate, and localized phase, respectively.
    }
    \label{fig:4}
\end{figure}

In Fig.~\ref{fig:4} we display in a single graph the system behavior as a function of the two relevant parameters, the potential strength $K$ and the degree of incommensurability $\Delta$. This plot, derived from the central-particle variance $\sigma_0^2$, can be considered a phase diagram that summarizes the general findings discussed so far. Larger incommensuration leads to larger values of the critical threshold $K_c$ above which the delocalized phase (blue region) is progressively lost, and the localized phase (red region) sets in. The size of the transition region (yellow region) progressively widens in the limit of large $K$ and $\Delta$ values, where incommensuration effectively fights the localizing effect imposed by the periodic optical potential. 

A similar, though less pronounced, widening is observed for small incommensurations. We infer that this might be due to the tunneling processes, that are effective for $K\simeq K_c$ at all incommensuration values. If we had in Fig. \ref{fig:4} considered instead the Binder cumulant as the indicator for the transition, the width in the low $K$-$\Delta$ region would be greatly reduced, indicating that the variance tends to overestimate the quantum effects. 
In essence, both incommensuration and quantum effects act to increase the fluctuations and resist localization, as intuitively expected. \\

\subsection{Spectroscopic observables}

Up to this point, we have characterized the transition for the whole system via the behavior of the central particle. As already discussed, on physical grounds this is exactly what happens in the fully commensurate case, where the springs can be ignored. However, we have already seen that with increasing incommensuration the transition becomes less sharp. In this and the next sections, we quantitatively trace the origin of this behavior to the collective properties of the system, and to the behavior of the other particles. We first consider the evolution of the total energy of the system, that, being a global quantity, should contain signatures of the behavior of all particles. 

\begin{figure}
    \centering
   \includegraphics[width =  1.\linewidth]{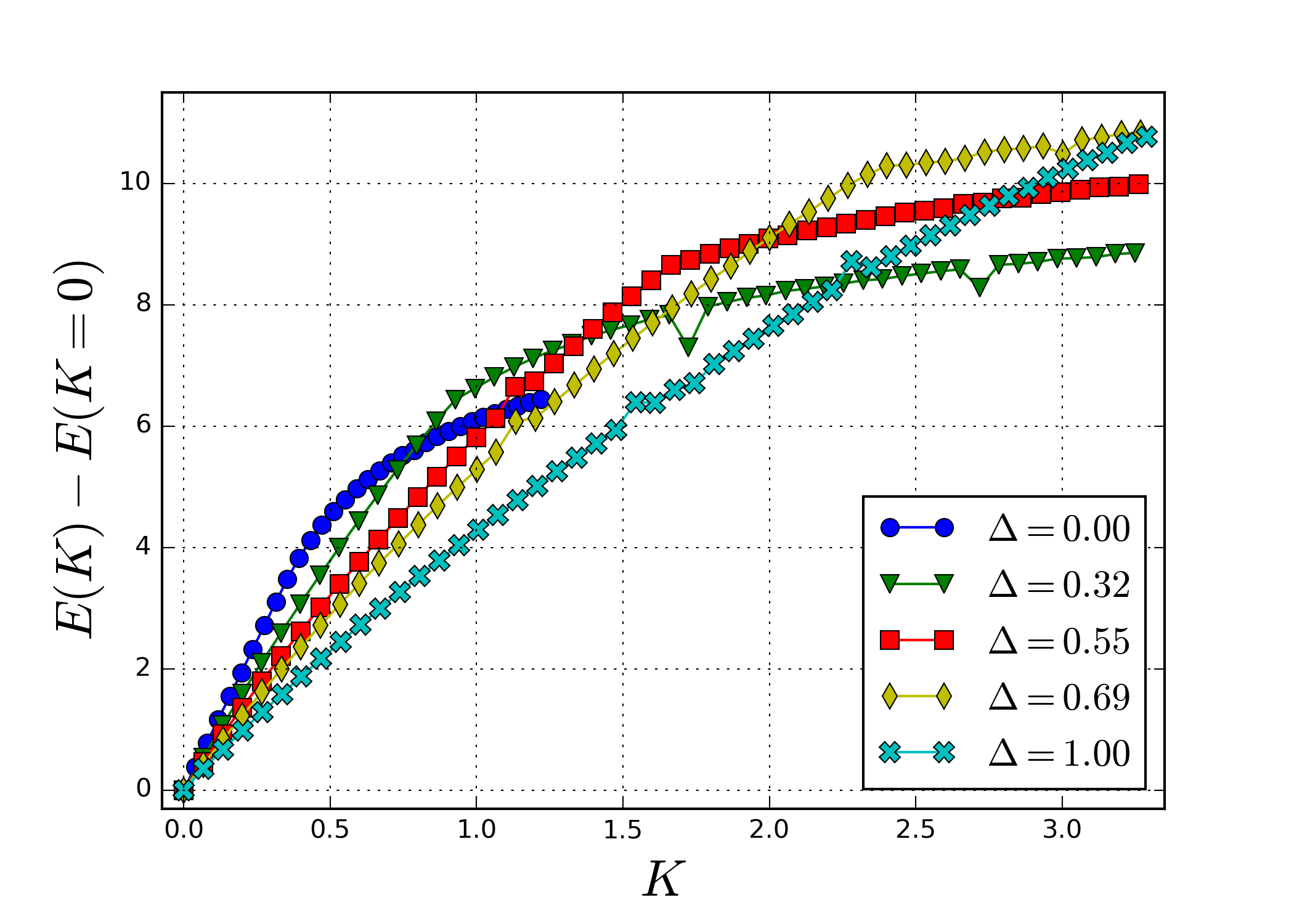}
    \caption{Behaviour of the system energy as a function of $K$ for different $\Delta$ values, measured with respect to the case with $K=0$. Once the transition occurs the energy growth-rate slows down, a sign that all the particles are localized in potential minima, so that the energy does not sensitively depend on $K$. Similarly to the variance in bottom Fig. \ref{fig3}, for the maximally incommensurate case the decrease in slope is never reached and the steepest behaviour occurs for intermediate incommensurations.}
    \label{fig5}
\end{figure}

The transition is more evident if instead of displaying the total energy, we consider the difference in energy between the lattice potential being switched on ($K\neq 0$) and off ($K=0$). (Experimentally, this can be measured directly by transferring the ions to a different internal state where they do not experience the light shift of the optical lattice.)
In this quantity also, a remarkable change occurs around the same critical $K_c$ as the one inferred from the maximum variance and the BC analysis. This is displayed in Fig.~\ref{fig5}, where the change is seen as a 'knee' in the curve, i.e. a change in the slope of the growth rate of differential energy with lattice potential K. This behavior can be understood with the following argument: When $K$ is increased in the sliding phase below $K_c$, the particles tend to be pushed closer to each other while trying to localize in the minima of the optical lattice. During this process, the energy must increase as a result of the interparticle elastic forces. No appreciable differences are encountered in this region between the quantum model and the calculation for the classical system (see SM \cite{SoM}). On the other hand, after the transition has occurred, and the particles begin to tunnel between different minima, the elastic energy grows less quickly in the quantum model. We finally remark that the energy difference in Fig.~\ref{fig5} and  the maximum variance in bottom Fig.~\ref{fig3} share similar behavior in the maximally-incommensurate regime with $\Delta=1$, where no transition is observed, even after extending simulations to considerably larger values of $K$.

\subsection{Hull function}
A second quantity that reflects the behavior of all particles is the hull function that is often used to characterize the incommensurate and commensurate phases~\cite{Hu2000,Bylinskii2016}. In our quantum system, this is the average position $\langle x_0 \rangle$ of the central particle as a function of the phase offset $\Phi$ between the periodic potential and the harmonic trap potential, see Eq. \ref{eq:H}. Here and in the following, the averaging refers to weighting with the quantum-mechanical probability density $\big|\Psi\big(x_{-({N-1})/{2}},...,x_{({N-1})/{2}}\big)\big|^2$, as determined from the $N$-ion wavefunction $\Psi$. 

\begin{figure}
    \centering
  \includegraphics[width =  1.\linewidth]{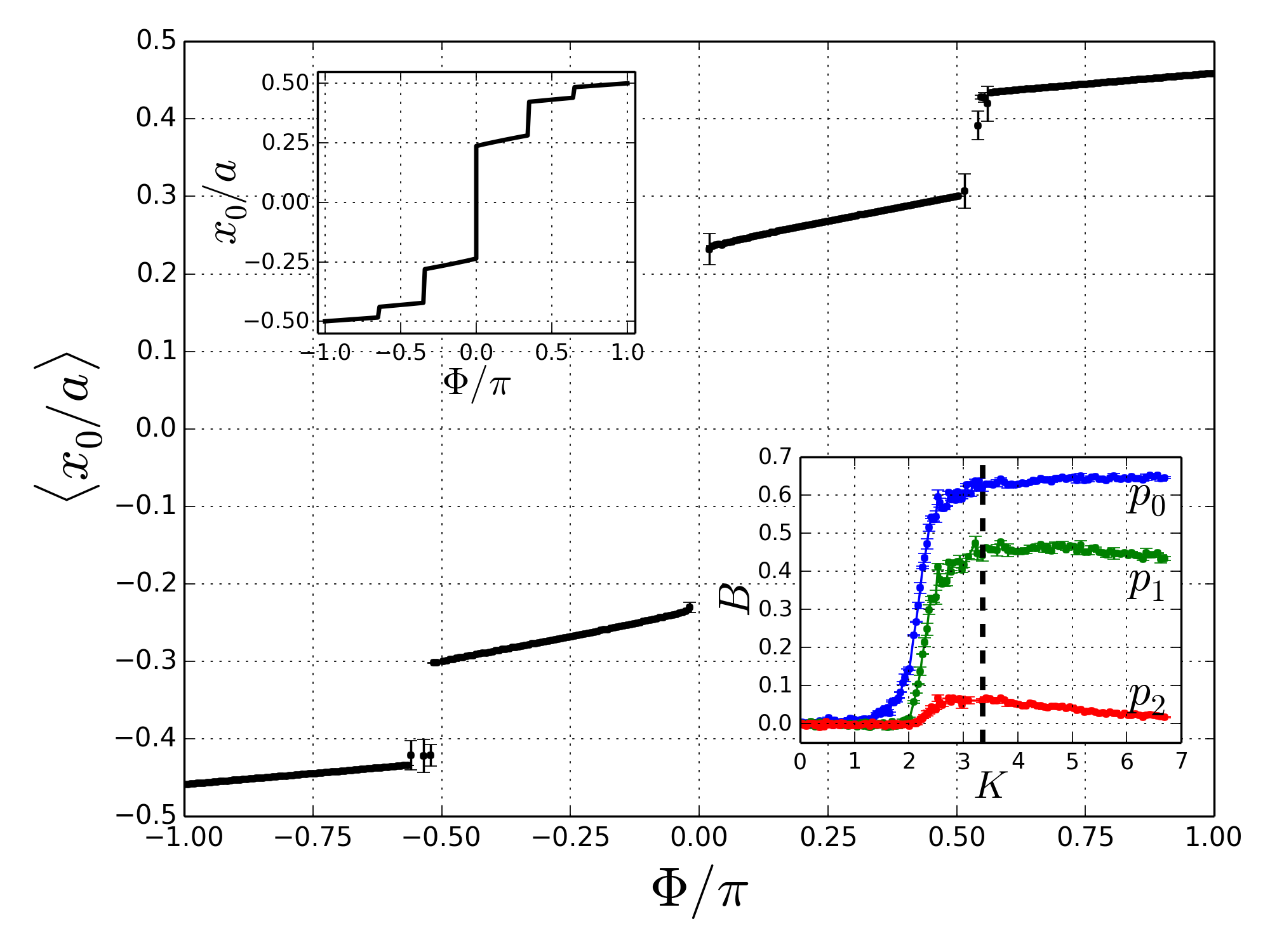}
    \caption{Hull function, i.e. the central particle average position as a function of the phase shift $\Phi$ for the maximally incommensurate system and $K=3.34>K_c$ above the pinning transition. Gaps are seen to open at each value of $\Phi$ for which one of the particles has to overcome a maximum of the optical potential. For $K<K_c$, the hull function would be a continuous function. Top-left inset: the Hull function in the classical case for the same parameters set as in the main figure. Notice that here all the gaps open up, while in the quantum case the tertiary gaps disappear. Bottom-right inset: Binder functions for the first three particles $p_{0,1,2}$ as a function of $K$: at $K=3.34$ as in the main figure (dashed vertical line) particle 2 is delocalized while particle 0 is localized.}
    \label{fig6}
\end{figure}

Below a threshold ($K<K_c$), the particles are  approximately uniformly arranged, and float on top of the lattice potential. The hull function in this region is continuous and monotonically increasing. In this characterization, the classical Aubry transition appears as a breaking of the analyticity of the hull function. Indeed we find that also for the finite quantum system, for $K > K_c$, as displayed in Fig.~\ref{fig6}, gaps in the hull function open up at each value of translation $\Phi$ for which one of the particles would be located at a local maximum of the optical potential. The appearance of multiple gaps signals the occurrence of incommensurations, where the particles no longer slip synchronously. We notice, however, that here only two out of the three possible gaps open up, and that the tertiary gaps disappear due to lack of localization of the particles at the edges of the chain. This is evident from the bottom-right inset, displaying the BCs for the first three particles (labelled as $p_{2},p_{1},p_0$ starting from the edge of the chain), as a function of $K$: at $K=3.34$ as in the main figure, particle 2 is delocalized while particle 0 is still localized. This behavior can be contrasted with the classical case displayed for the very same set of parameters in the top-left inset of the figure, where all the three gaps are seen to open up. The corresponding behavior has also been observed experimentally (see Figs. 1e and 1f in Ref.~\cite{Bylinskii2016}).

\subsection{Correlations}
\begin{figure}[!h]
\centering
{\includegraphics[width=1.\linewidth]{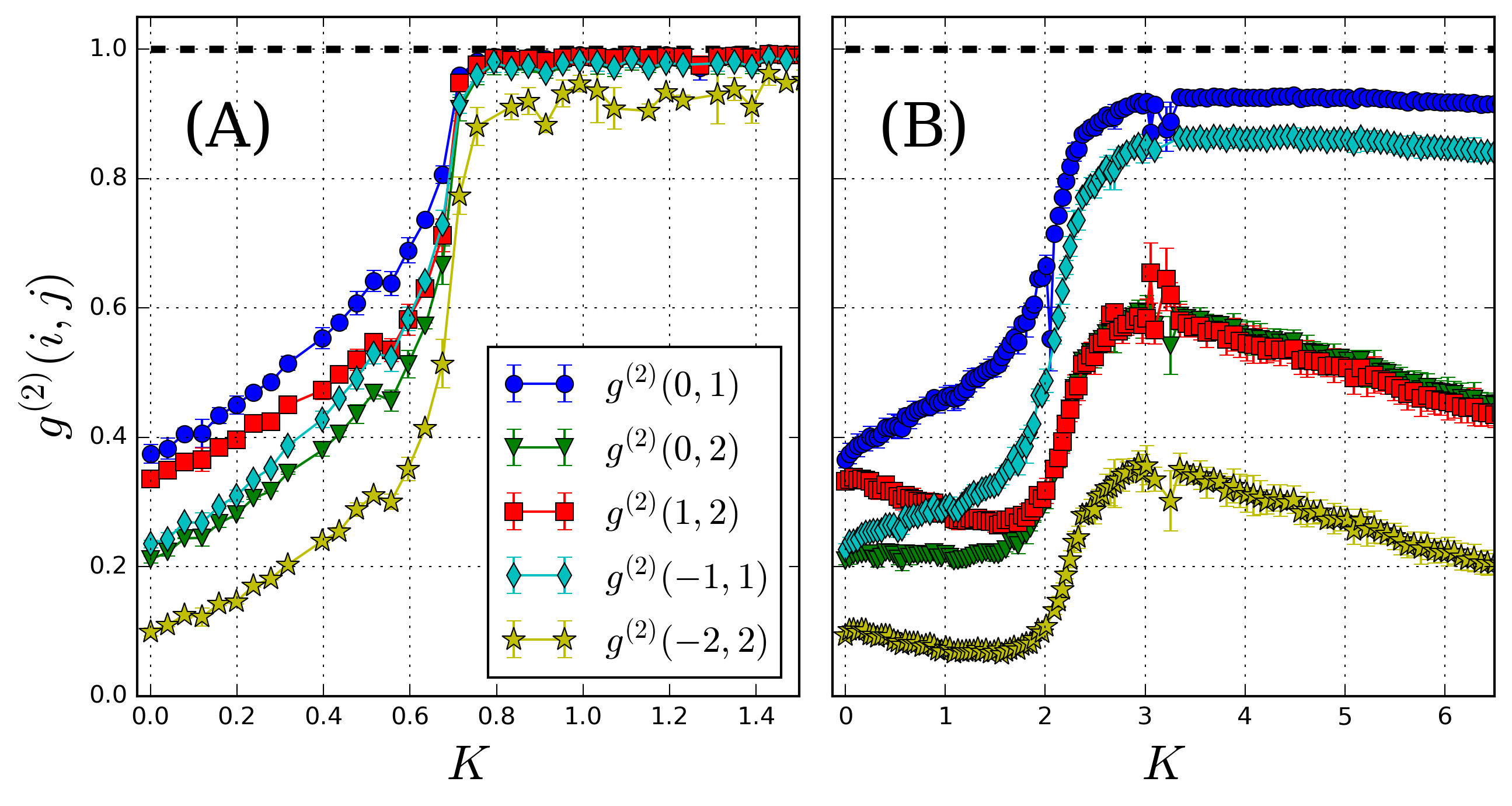}} 
\caption{Many-particles effects. Two-particle correlations as a function of $K$. The particles are labelled in a way that 0 stands for the central particle, 1 for its right first neighbour, -1 for its left first neighbour, \textit{etc.}. (A) (Almost) maximally commensurate case. (B) (Almost) maximally incommensurate case. The line $g^{(2)}(i,j)=1$ at which two particles are maximally correlated is reported in both figures.}
\label{fig7}
\end{figure}

We now explicitly look at the other particles behavior by analyzing suited two-particle correlations vs. $K$. These are defined as
\begin{equation}
g^{(2)}(i,j)=\frac{\langle (x_i-\overline{x}_i)(x_j-\overline{x}_j)\rangle}{\sqrt{\sigma^2_i \sigma^2_j}}.   
\end{equation}
Fig.~\ref{fig7} illustrates the behavior of $g^{(2)}(i,j)$ in the two extreme cases of a maximally commensurate (left panel) and of a maximally incommensurate (right panel) system.
In the commensurate case on the left panel,  
above a critical $K$ value all the correlations saturate to 1, signalling a highly coherent behaviour. We notice that, when compared to the commensurate case, all the pair correlations in the incommensurate regime are smoother and never saturate to 1, a strong sign that while the central particle undergoes the pinning transition, the particles at the edges remain delocalized. 

The correlation pattern between the particles is non-trivial. For example, correlations between the (1,2) and the (-1,1) pairs cross each other as a funtion of potential depth $K$. For small $K$, the $(1,2)$ correlation is larger because it involves two neighbours, while the $(-1,1)$ is a weaker second-neighbour correlation. However, for large $K$ the $-1$ and $1$ particles undergo the pinning transition, and become highly correlated, while the outer particles $2,-2$ remain delocalized. We finally notice that, similarly to what happens to the BCs and the maximum variances, the correlations involving particles $1,-1,2$ and $-2$ decrease for large $K$ values.

\subsection{Temperature and size effects}
Finally, we verify to what extent the observed behavior across the transition is influenced by finite temperature or the finite size of the system. Regarding the role of temperature, we have compared simulations for two temperatures differing by one order of magnitude, with all other parameters kept fixed. As detailed in the SM~\cite{SoM}, no significant differences have been found in the three indicators of the transition, the BC, maximum-variance, and the system energy. We thus conclude that quantum effects can be observed even above the lowest accessible temperature.

Second, the harmonic confinement favors delocalization of the most external particles, and thus one might ask whether the effects found in the calculations would persist in a homogeneous system. In order to mimic the system in the absence of external confinement under stable conditions, we have performed the same simulations in a ring geometry with the harmonic confinement switched off. As detailed in the SM~\cite{SoM}, the resulting BC and maximum variance indicators provide a consistent and similar message as for the inhomogeneous system. In addition, we observe the persistence of the decaying correlation-functions behavior shown in Fig.~\ref{fig7} as well as the energy behavior in Fig.~\ref{fig5}.  

\section{Prospects for the experimental observation of quantum effects in the Aubry transition} \label{sec:Experiment}

The parameters used for the calculations described above are in the same range as in the experiment of Bylinskii et al \cite{Bylinskii2015}. In particular, the temperatures where quantum effects are observable have already been reached in the experiment using Raman sideband cooling \cite{Karpa2013}, and even lower temperatures can be reached using cooling on a narrow optical-clock transition \cite{Taylor1997}. Therefore finite temperature should not constitute an obstacle that would prevent the observation of a quantum Aubry transition.

Among the quantum signatures of the Aubry transition discussed in the current work, the energy of the system is the most directly accessible. It can, e.g., be directly measured spectroscopically by probing a transition to an internal state in which the ions do not experience the periodic optical potential. For instance, in the experiment by Bylinskii et al.  \cite{Bylinskii2015}, the periodic potential was generated by an optical lattice relatively near detuned to a transition $^2S_{1/2} \rightarrow ^2P_{1/2}$. This optical lattice creates a periodic potential for the electronic ground state $^2S_{1/2}$, but for instance not for the long-lived electronic excited states (clock state \cite{Taylor1997}) $^2D_{5/2}$. Thus with one laser beam addressing all ions, one could directly probe the energy difference between the system with optical potential (in the ground state $^2S_{1/2}$) to the system without optical potential (in the clock state $^2D_{3/2}$). This would enable one to map out a curve as shown in Fig. \ref{fig5}, and verify the spectroscopic signature of the transition. Alternatively, one can also probe excitations of the system by optically addressing only a single ion and transferring it to the state $^2D_{3/2}$ where it does not experience the periodic potential.

The optical resolution with which the atom position can be directly measured, on the other hand, is typically insufficient to determine the BC or the change of the variance of the atomic position across the Aubry transition. However, as demonstrated in the experiments \cite{Bylinskii2015,Gangloff2015,Bylinskii2016}, the atomic fluorescence during the continuously applied laser cooling depends sensitively on the position of the atom relative to the optical lattice. Using this tool, it has been possible to determine the average positions of all the ions across the transition, with resolution well below the lattice period $a$ \cite{Bylinskii2015}. Extending this technique, it may be possible to use the statistics of the continuously detected atomic fluorescence light to infer the position distribution for each atom, and to reconstruct the BC and the variances of the atomic positions.

\section{Conclusions}\label{sec:Conclusions}

In conclusion, we have performed PIMC simulations precisely reproducing the conditions of the experiment on an Aubry-type transition in a finite ion chain~\cite{Bylinskii2016}. We have found that quantum effects in the Aubry transition emerge in two different observables never considered before as indicators, namely, the Binder cumulant, and a suitably devised spectroscopic technique measuring the difference in energy between the system with and without exposure to the optical lattice. The maximum variance may be added as a third signature of the transition. For all three observables, a steep, sudden change of behavior consistently occurs at a given critical value $V_c$ depending on the incommensuration parameter $\Delta$ as summarized in Fig.~\ref{fig:4}. At least some of these quantities can be measured in the experiment \cite{Bylinskii2015}, leading to a promising route to identifying the impact of quantum effects on the Aubry transition. Even more remarkably, the transition appears to be robust against finite-size effects, and may be discernible even in relatively small chains of ions. It would be interesting to follow the evolution of the transition with increasing ion number, although the experiment would eventually be limited by stability conditions, while our particular PIMC is limited by increasing simulation times. Refs. ~\cite{Hu2000} and ~\cite{Benassi11} comment that the transition appears to be second order. However, we feel that a definite answer to this question can only be given after identifying a suitable order parameter, as none of those considered in literature is fully suitable.

The focus of the present work is on the observability of quantum effects, a non-trivial task in the presence of Coulomb long-range interactions~\cite{Retzker2008,MITUNIPI}, and augmenting existing work on the quantum Aubry transition \cite{Benassi11,Hu2000}. It paves the way for studies along a number of different directions to be explored in future combined theoretical-experimental effort~\cite{MITUNIPI}. This includes the possibility of elucidating the universality class of the transition while tuning the degree of quantumness, or to assess whether and to which extent the transition can be related to the paradigm of many-body localization. Finally, the very general nature of the Aubry transition, as occurring in so many diverse classical and quantum systems, can in principle be probed against microscopic mechanisms. This would be a natural operation to be performed, e.g., using available platforms of ultracold atomic quantum gases. In fact, we may envisage a number of different microscopic mechanisms. For example, using ions in optical cavities, one may build  Ref.~\cite{Cormick} and investigate the robustness of mechanical stability against quantum dynamics, or else engineer the quantum reservoir. The cases of dipolar and contact-like interactions can be especially interesting in shining light on whether and how the interaction range affects the transition, with Rydberg-atom systems covering the intermediate interaction range. Finally, by controlling Rydberg atoms inside optical cavities, one can access an intriguing scenario where the infinite-range nature of the effective interatomic interaction mediated by cavity photons is combined with the strong and coherent medium-range interaction between Rydberg atoms.

\begin{acknowledgments}
This work has been supported by the NSF, the NSF-funded Center for Ultracold Atoms, and by the MIT-UniPi Program. We thank Giovanna Morigi for very fruitful discussions. MLC and PMB are grateful to MIT for hospitality.  
\end{acknowledgments}

%
\end{document}